# Crowdsourced Smartphone Sensing for Localization in Metro Trains

Haibo Ye
State Key Laboratory for Novel Software Technology
Nanjing University, Nanjing, China
yhb@smail.nju.edu.cn

Tao Gu
School of Computer Science and IT
RMIT University
Melbourne, Australia
tao.gu@rmit.edu.au

Xianping Tao, Jian Lu
State Key Laboratory for Novel Software Technology
Nanjing University, Nanjing, China
{txp, lj} @nju.edu.cn

***Abstract*—Traditional fingerprint based localization techniques mainly rely on infrastructure support such as RFID, Wi-Fi or GPS. They operate by war-driving the entire space which is both time-consuming and labor-intensive. In this paper, we present M-Loc, a novel infrastructure-free localization system to locate mobile users in a metro line. It does not rely on any Wi-Fi infrastructure, and does not need to war-drive the metro line. Leveraging crowdsourcing, we collect accelerometer, magnetometer and barometer readings on smartphones, and analyze these sensor data to extract patterns. Through advanced data manipulating techniques, we build the pattern map for the entire metro line, which can then be used for localization. We conduct field studies to demonstrate the accuracy, scalability, and robustness of M-Loc. The results of our field studies in 3 metro lines with 55 stations show that M-Loc achieves an accuracy of 93% when travelling 3 stations, 98% when travelling 5 stations.**

## I. Introduction

With a growing number of mobile phones, Location Based Services (LBS) have become more and more popular. Finding mobile user's location (i.e., localization) is one of the key enabling technologies. An outdoor navigation service typically use GPS-enabled mobile phones to find nearby places. There has been an increasing demand for LBSs used in indoor environments. The Google Maps [1] for mobile phones will support indoor navigation [2] which shows where you are and guide you to where you want to go in an indoor environment. For example, in a shopping mall scenario, this service provides mobile users with turn-by-turn navigation to where they want to go. In such an indoor environment where GPS is not available, many Wi-Fi based localization techniques [4-6] [8] have been proposed and widely used. For example, RADAR [6] uses Wi-Fi signal strength. The idea is to war-drive a building, and then create a radio map linking physical locations and Wi-Fi fingerprints.

While the main focus of indoor localization research has been centered around in-building scenarios, no attention is given in locating mobile users when they are taking metro[1] trains. Metro trains have been the most important means for urban transport, and every large city in the world has metro systems in operation. Localizing mobile users in metro lines not only extends existing LBSs to this blind spot, but also promotes and encourages the use of public transport in achieving urban sustainability.

How to locate mobile users in a metro train is not a trivial task. GPS does not work underground. A Wi-Fi fingerprint based approach may be applied in this scenario provided Wi-Fi access points can be deployed along each metro line. However, Wi-Fi based solutions obviously not only incur high installation and maintenance costs, but also increase the chance of interfering with the metro control system. Today, only a few metro lines provide Wi-Fi coverage, mostly in metro stations only. Until 25 Apr 2013, in New York, there are only 36 metro stations have Wi-Fi access (i.e., 13% of all metro stations [3]). Hong Kong has just start providing Wi-Fi access from 8 July 2013 for a limited number of metro stations. Some metro control systems (e.g., Siemens' EMCS system [18]) use Wi-Fi to keep metro trains in connection with the control center, but they are not public accessible. In addition, when applying Wi-Fi fingerprinting in metro lines, war driving is not an easy task since it is difficult to obtain Wi-Fi fingerprints when trains are running underground.

What if we don't use Wi-Fi? An intuitive way is to use RFID techniques like [19] to locate mobile users in a metro line. This requires that every mobile user carry a tag, and RFID readers are deployed in each metro cabin. When a user enters into a train, the reader gets the tag ID, and the location of the train can be obtained from the metro control system. This approach needs infrastructure support which is costly, and it leverages the train location system.

Motivated by the recent advance of smartphone sensing, in this paper, we propose to use magnetometer and barometer sensors on smartphones and build sensor fingerprints through crowdsourcing which requires neither infrastructure nor war driving. Our approach is based on the following observations. When a train is running in the tunnel from station A to station B, the magnetometer and barometer readings scanned from a user's smartphone show unique patterns due to a variety of physical environments in different tunnels. Such patterns can be exploited as fingerprints to locate the train by pattern matching. To get the fingerprints of magnetic field and barometric pressure in each tunnel, different from the traditional approach which war drives the entire metro line, we leverage on crowdsourcing. Each mobile user taking the metro will contribute partial data which will be then combined for extracting patterns by an algorithm. To achieve this, we first design a scalable, noise-free event detection algorithm to detect the event of train stopping or leaving a station based on accelerometer and barometer readings. By knowing these events, we can then obtain the barometer and magnetometer readings when a metro train is running in a tunnel. Second, the readings are collected and uploaded into a cloud server which runs our DTW [16] based pattern matching algorithm to extract and compare the patterns for different tunnels. We then merge different user traces by a merging algorithm to generate a pattern graph for the tunnels. By comparing the graph with the metro line map, we are able to link each pattern in the graph to a specific tunnel. A mobile user can then download the pattern map from the cloud sever for localization by looking up the map. In summary, we make the following contributions:

1) We propose a novel metro train localization approach, named M-Loc, to identify the locations of mobile users in metro trains. Compared to traditional Wi-Fi fingerprinting based approaches, M-Loc requires neither infrastructure nor war driving.

2) Using smartphone sensing and crowdsourcing, we propose several algorithms to handle noisy sensor readings and perform pattern matching, which make our solution more practical and scalable.

3) We conduct a field study in a real metro system with 3 metro lines and 55 stations, and analyze the performance of M-Loc. The results show that M-Loc achieves an accuracy of above 90% when a passenger travels 3 stations and over 98% when travels 5 stations.

[1] Metro, also named subway or underground, in this paper, refers to the trains run underground.

The rest of this paper is organized as follows. Section 2 gives the overview. The detailed design is shown in Section 3. Section 4 describes our evaluation. Section 5 discusses the related work, and finally, Section 6 concludes the paper.

## II. System Overview

We give an overview of M-Loc in this section, as shown in Fig. 1. The system operates in two phases. In the first phase, we crowdsource both magnetometer and barometer readings from passengers' smartphones. We then analyze the patterns from these readings to extract the unique pattern for each tunnel (i.e., a tunnel between two adjacent stations), and generate the pattern map. In the second phase, users download the map for localization.

### A. In the first phase

M-Loc builds the pattern map of a metro line by crowdsourcing. In order to get patterns of all the tunnels in a metro line, our idea is to first extract patterns from user traces, and then discover the patterns which are linked to specific tunnels. With crowdsourcing, each user contributes a complete or partial trace of his entire trip when travelling in a metro line. We detect the event of train stopping or leaving a station, and collect both magnetometer and barometer readings when the train is running in the tunnels. Each user trace contains only partial patterns of a metro line. By merging the traces from different users, we are able to obtain a complete graph which contains all the patterns of the metro line. The structure of the graph can then be mapped to the real metro line map, and links each pattern to a specific tunnel.

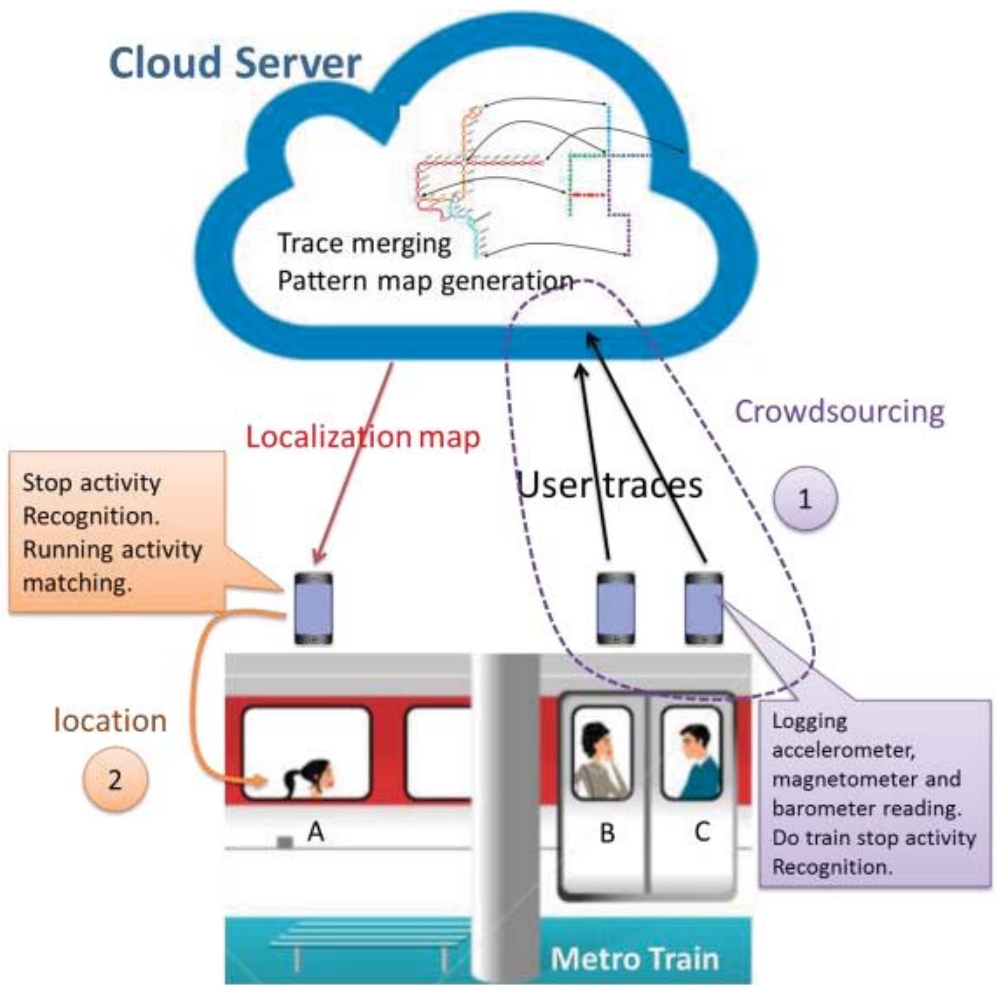

Fig. 1. The overview of M-Loc

In more details, when a user travels in a metro train, the mobile client software collects acceleration, barometric pressure and magnetic field data. To recognize the event of train stopping at a station, we use acceleration and barometric pressure data. The acceleration data show clear signatures when the train decelerates to stop and accelerates to leave. In addition, the barometric pressure in the train will show clear jumps when the doors of the train open or close. Combining these two signatures, we are able to accurately detect train stopping at a station, and we name it a *train stop event*. Meanwhile, we name the process when the train is running in the tunnel between two adjacent stations a *train running event*. By these train stop events, we divide a stream of barometer and magnetometer readings arrived in time order into fractions, and each fraction associated with either a train stop event or a train running event. We call the sequence of stop and running events as a user trace. When a user leaves the metro train, the trace will be uploaded to the cloud server. In the cloud server, we know each running event occurs in a tunnel, but which tunnel is unknown. Since the train running events in each tunnel have a common unique pattern, we design a DTW based pattern matching algorithm to find all the running events of the same tunnel. We then apply a merge algorithm to merge the traces from different users. Finally, we generate a pattern graph which covers the entire metro line. We can map the running event patterns to the tunnels by comparing the pattern graph with the metro line map, and get the patterns of all the tunnels.

### B. In the second phase

For localization, users download the pattern map from the cloud server to their smartphones. When a user travels in a metro train, both barometric pressure and magnetic field data are logged, and the train stop and running events are detected in real time. The pattern matching algorithm will be used to match the train running event to the tunnel based on the pattern map. Finally, the location of the train is known, so is the location of the user.

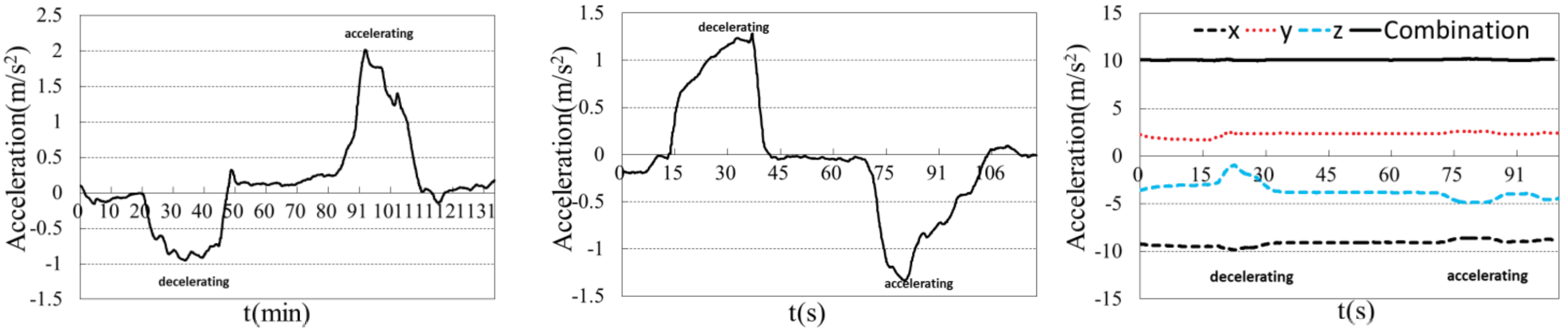

(a) The x-axis acceleration readings when the x-axis of the phone is pointing to the front of the train

(b) The x-axis acceleration readings when the x-axis of the phone is pointing to the back of the train

(c) The acceleration readings when the phone is randomly placed

Fig. 2. The acceleration readings when a train stops and leaves a station

## III. SYSTEM DESIGN

We first give the details of the train stop event detection and the pattern map generation, we then describe how to use the map for localization.

### *A. Train Stop Event Detection*

In order to detect a train stop event, we first analyze the available sensors on smartphones, and justify our choice of accelerometer and barometer. Later, we show the detailed approach for detection.

#### *1) Accelerometer signature*

A metro train stops at a station follows a similar process − decelerating to stop at a station, open the doors, close the doors and accelerating to leave. It is clear that accelerometer can capture these motion events. The readings of the accelerometer will show a clear signature. As shown in Fig. 2(a), it appears as a crest when decelerating and a trough when accelerating. Actually in the real case, the acceleration signature may show slightly different depending on the direction of the smartphone. Fig. 2(a) shows the signature when the x-axis of the smartphone is in the same direction of the train, while in Fig. 2(b) the x-axis is in the opposite direction of the train, where it appears as a trough when decelerating and a crest when accelerating. What appears most often is the situation where a phone's direction is random, as shown in Fig. 2(c), the acceleration of the train when accelerating and decelerating appears in the three axes of the accelerometer, each with a smaller crest or trough.

#### *2) Barometer signature*

In addition to accelerometer, we observe that barometer of smartphone also shows a clear signature in the process of a train stop event. With air conditioning or ventilation equipment used in every train, barometric pressure appears different between the inside and outside of a train cabin. For this reason, when the train stops and opens the doors, there exists a sharp drop in barometric pressure. On the contrary, there exists a sharp increase in barometric pressure when the doors close. This change appears clearly in the readings from the barometer sensor. We show this signature in Fig. 3(a) and it is an important feature for detecting a train stop event.

#### *3) Issues with acceleration*

Each acceleration sample is a triple, including a reading of x, y and z, respectively, each represents a direction of the smartphone, as shown in Fig. 3(b)(1). It should be noted that the acceleration caused by gravity is included in the triple. When the train is accelerating or decelerating, the readings are the combination of the acceleration of the train and the gravity; this is shown in Fig. 3(b). We observe a sharp crest and trough from Fig. 2(a) for only one axis, but a weak crest and trough from Fig. 2(c) since it distributes among all the three axes. These observations are due to various smartphone's orientations. To detect the train stop event, we need to find the pair of crest and trough. Assume that we use only one axis data. For the case in Fig. 2(a), the crest and though are clear and easy to detect. For the case in Fig. 2(c), every crest or though is not clear and is not easy to detect. We combine the readings from all the three axes. Fig. 2(c) shows the result of combining three axes. As we can see, both the crest and trough almost disappeared. This may be due to gravity, and the direction of accelerating/decelerating and the direction of gravity are mutually perpendicular. A slight acceleration change in the vertical direction of gravity will cause very small change to the combination of the acceleration. For example, the gravity is $9.8 m/s^2$, when the train accelerating is about $2\ m/s^2$, the combination is changed from $9.8\ m/s^2$ to $10\ m/s^2$, only $0.2\ m/s^2$ change of accelerometer reading is not obvious enough for detecting a train stop event.

#### *4) Using variation of acceleration*

To solve this issue, we use the variation of acceleration. Since the gravity keeps unchanged, the variation of acceleration is only affected by train acceleration. The combined variation of the three axes is the variation of train acceleration. The way to get the variation of each axis and the combination are shown in Equations 1-7. The direction of the combination is set as the direction of the axis with the max mean variation. Fig. 3(c) shows the combination of the variation of the three axes. In this figure, the deceleration process is transformed to a curve ending with a sharp crest. Similarly, the acceleration process starts with a sharp crest. By this transformation, the variation will not be affected by the phone orientation and the gravity.

$$\mathrm{VarX_a(t) = X_a(t+\Delta t) - X_a(t),\ VarX_{all} = \sum|VarX_a(t)|} \quad (1)$$

$$\mathrm{VarY_a(t) = Y_a(t+\Delta t) - Y_a(t),\ VarY_{all} = \sum|VarY_a(t)|} \quad (2)$$

$$\mathrm{VarZ_a(t) = Z_a(t+\Delta t) - Z_a(t),\ VarZ_{all} = \sum|VarZ_a(t)|} \quad (3)$$

$$\mathrm{Var_{max} = Max\,\{VarX_{all},\ VarY_{all},\ VarZ_{all}\}} \quad (4)$$

$$\mathrm{VarK_a(t)} = \begin{cases} \mathrm{VarX_a(t)\ \ if\,Var_{max} = VarX_{all};} \\ \mathrm{VarY_a(t)\ \ if\,Var_{max} = VarY_{all};} \\ \mathrm{VarZ_a(t)\ \ if\,Var_{max} = VarZ_{all};} \end{cases} \quad (5)$$

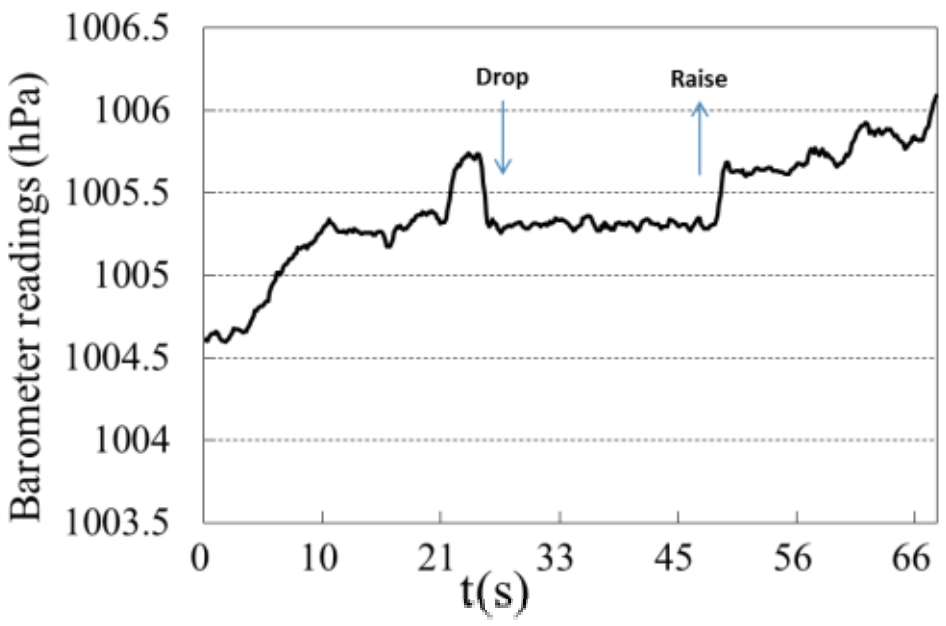

(a) The barometer readings signature when the train's door open and close

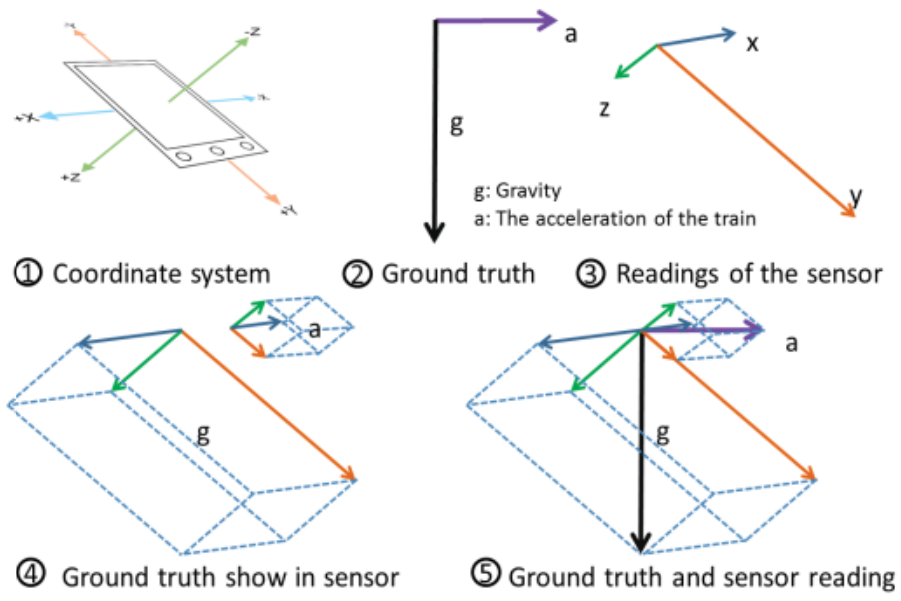

(b) The relation of the acceleration of the train and gravity

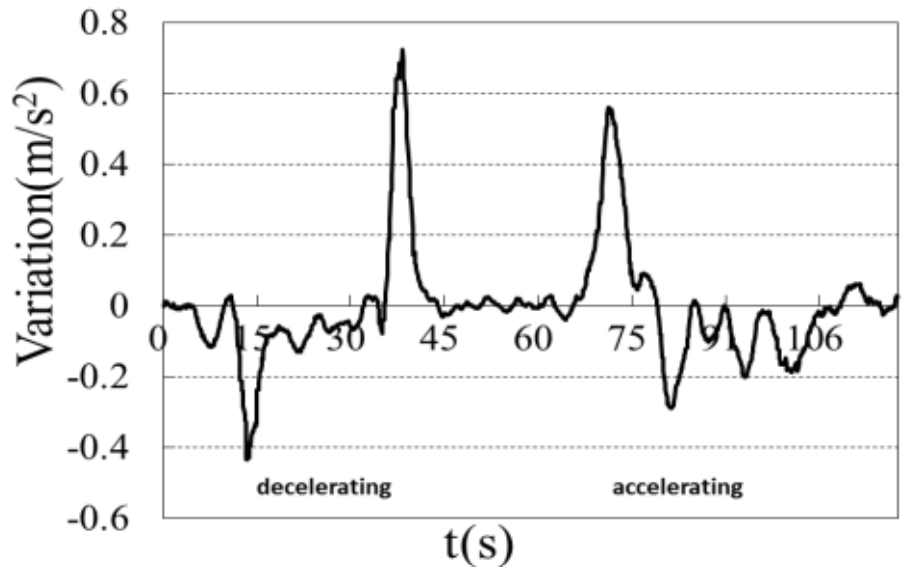

(c) The combination of the accelerometer readings variation in three axis

Fig. 3. The sensor readings of the smartphone on the train

$$Dir(t) = VarK_a(t) \,/\, |VarK_a(t)| \quad (6)$$

$$\mathrm{Var_All_a(t)} = \mathrm{Dir(t)} \times \sqrt{\mathrm{VarX_a(t)^2 + VarY_a(t)^2 + VarZ_a(t)^2}} \quad (7)$$

Where, $X_a(t)$ is the acceleration in x axis at time t. $\mathrm{Var_All_a}(t)$ is the combination variation of three axes at time t.

To detect a train stop event, we use a state machine shown in Fig. 4. A stop event includes the start and end time, it is defined as

$$\mathrm{StopEvent: SE = \{bt, et, ID\}},$$

where bt is the start time of the event, et is the end time of the event, and ID is the metro station where the event occurs. For the example in Fig. 3(c), the stop event is represented as $\{40,65,?\}$. At this stage, when a train stop event is detected, we get bt and et, but ID is unknown.

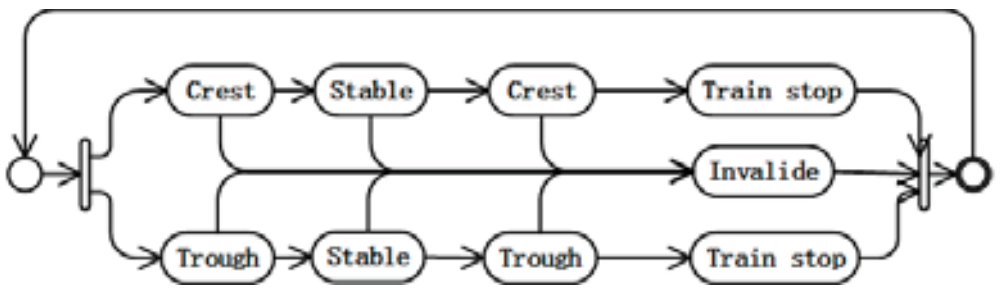

Crest and Trough: variance > 0.4, stable: variance < 0.1, stable time > 10 seconds.

Fig. 4. The state machine of detecting a stop event by acceleration.

### 5) *Enhancement with barometer*

The acceleration based approach may experience wrong detections and miss detections. We achieve an accuracy of about 85% based on our experiments. For example, when a train is running in a tunnel, the train driver may decelerate to control the speed for a short period and accelerate again. The acceleration readings show the same signature as a stop event in a station. This may happen especially during rush hours.

In order to filter out these false detections, we make use of barometer to detect an event of door opening or closing. For all the train stop events detected using acceleration readings, we check whether a door open or close event has occurred using barometer readings. In Fig. 3(a), the readings experience a sudden drop for about 0.4 hPa when the door opens, keep stable for some seconds and experience a sudden increase for about 0.4 hPa when the door is closed. We also use a state machine to detect this signature, which is shown in Fig. 5. We don't use the barometer to detect train stop only because it is not reliable in the following situations. For example, when the door opens and closes more than once in a station and when the barometric pressure drop and raise but not caused by door open and close.

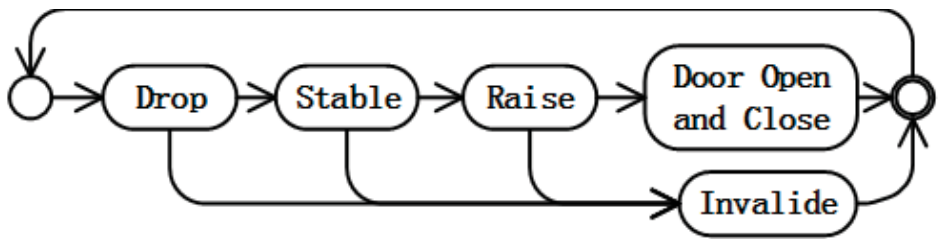

Drop and Raise: pressure_change > 0.3, stable: variance < 0.15, time > 10 seconds.

Fig. 5. The state machine to check a stop event by barometer readings.

### 6) *Accuracy of train stop event detection*

To test the accuracy of our approach, we hired 4 students who commute by metro train every day for experiments. Fig. 6(a) shows their routes. A data collection application runs in background on their Android smartphones to log sensor readings such as barometer, accelerometer, magnetometer and microphone. The experiment runs for a week. The audio recorded by microphone on smartphone will be used as the ground truth for which metro stations a user travels. We can easily obtain the ground truth such as the time of a train stops at a station by playing back the audio recorded. Comparing to result from our event detection algorithm, we obtain the accuracy as follows. Out of a total number of 427 train stops, 410 of them are successfully detected. The accuracy is about 96% with 3 wrong detections and 17 miss detections. If we only use acceleration readings, we get only 363 right detections. The comparison is shown in Fig. 7.

After detecting all the train stop events, we get data of the train running events. A running event is defined as:

$$\mathrm{RunningEvent: RE = \{bt, et, BTrace, MTrace, ID_1, ID_2\}}$$

where bt is the start time of the running event (i.e., the time when the train leaves a station), et is the end time of the running event (i.e., the time when the train arrives at the next station), and BTrace is the barometer trace scanned between time bt and et. A barometer sample is represented as $B = \langle t, baro\rangle$ and $BTrace = \langle B_1, B_2, \cdots\rangle$. MTrace is the magnetic field trace scanned between time bt and et. A magnetic field sample is $\mathrm{M} = \langle \mathrm{t, x, y, z, m}\rangle$, which is the magnetic field readings in the three axes of the smartphone at time t, and $\mathrm{m} = \sqrt{\mathrm{x^2 + y^2 + z^2}}$. $\mathrm{MTrace} = \langle \mathrm{M_1, M_2, \cdots}\rangle$, and M is in time order. $\mathrm{ID_1}$ and $\mathrm{ID_2}$ are the metro stations between which the train running event occurs, $\mathrm{ID_1}$ and $\mathrm{ID_2}$ also identify a tunnel.

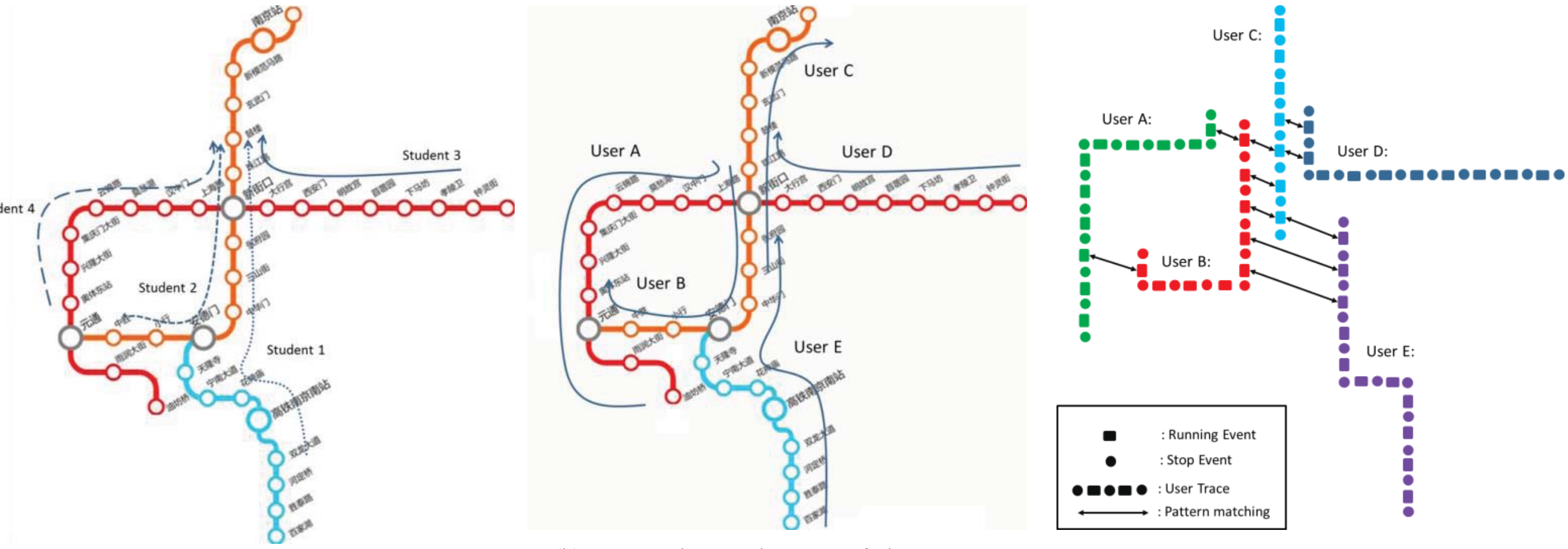

(a) The real travel trace of the four students

(b) Example travel traces of the users that cover the entire metro line

(c) User traces and the matching of running events

Fig. 6. User traces and matching

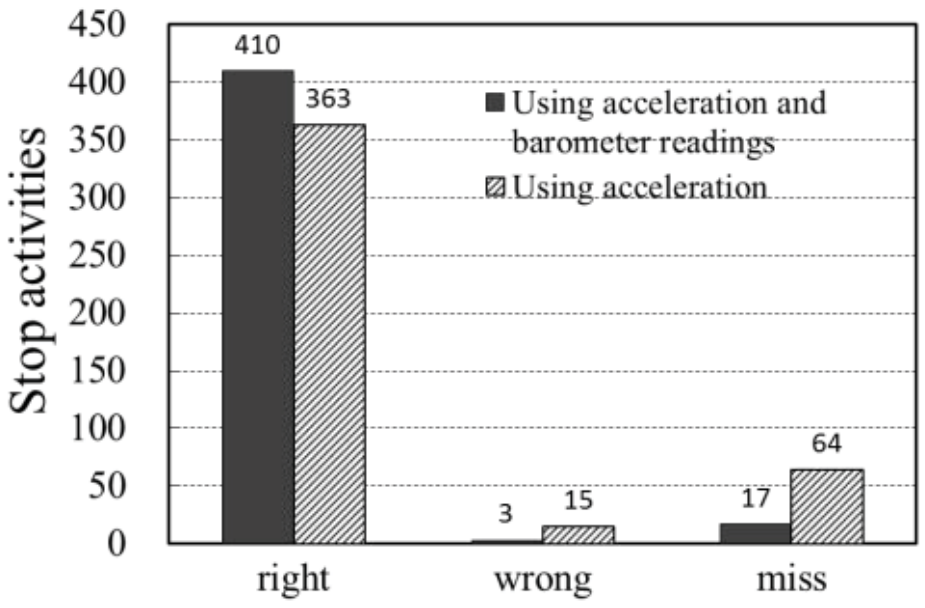

Fig. 7. The result of detecting train stop events

We have now obtained both barometric pressure and magnetic field patterns from a train running event, but we don't know in which tunnel the running event occurred. In the following section, we describe how to map a train running event to a tunnel.

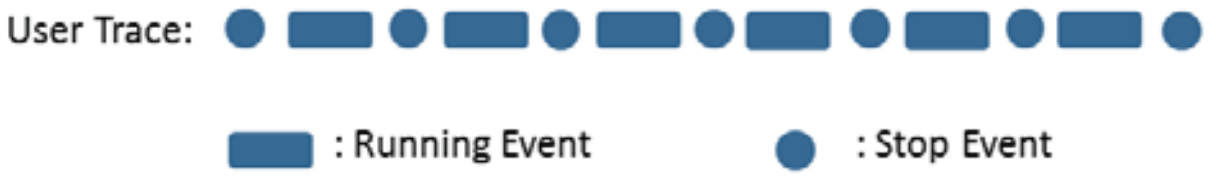

Fig. 8. A user trace of passing 7 train stations and 6 tunnels

## B. *Pattern map generation*

The pattern map contains all the patterns of the tunnels in a metro line, and each pattern is mapped to a specific tunnel. In each user's trace, there exist one or more train stop and running events. We now present our approach to build the pattern map.

We define a sequence of train stop and running events as a user trace, denoted as $\mathrm{UserTrace} = \langle \mathrm{SE}_1, \mathrm{RE}_1, \mathrm{SE}_2, \mathrm{RE}_2, \cdots \rangle$, where SE is a stop event and RE is a running event. Fig. 8 shows a typical trace of a user. We know that the running event occurs in a metro tunnel. With enough traces, the tunnels where the running events occur will cover all the tunnels in the metro line, on the condition that every metro station has been visited at least once. Furthermore, the traces from different users may have overlapped tunnels. An example is shown in Fig. 6(b), 5 users contribute 5 traces which eventually cover the entire metro line. Overlapped tunnels exist in their traces. We will get five discrete traces, but the tunnels of the traces are unknown. As shown in Fig. 6(c), we use pattern matching to merge the traces with overlapped running events and build a graph of running and stop events. Using the real map of the metro line which is public accessible, we can map the running and stop events to the tunnels and stations, which are shown in Fig. 9.

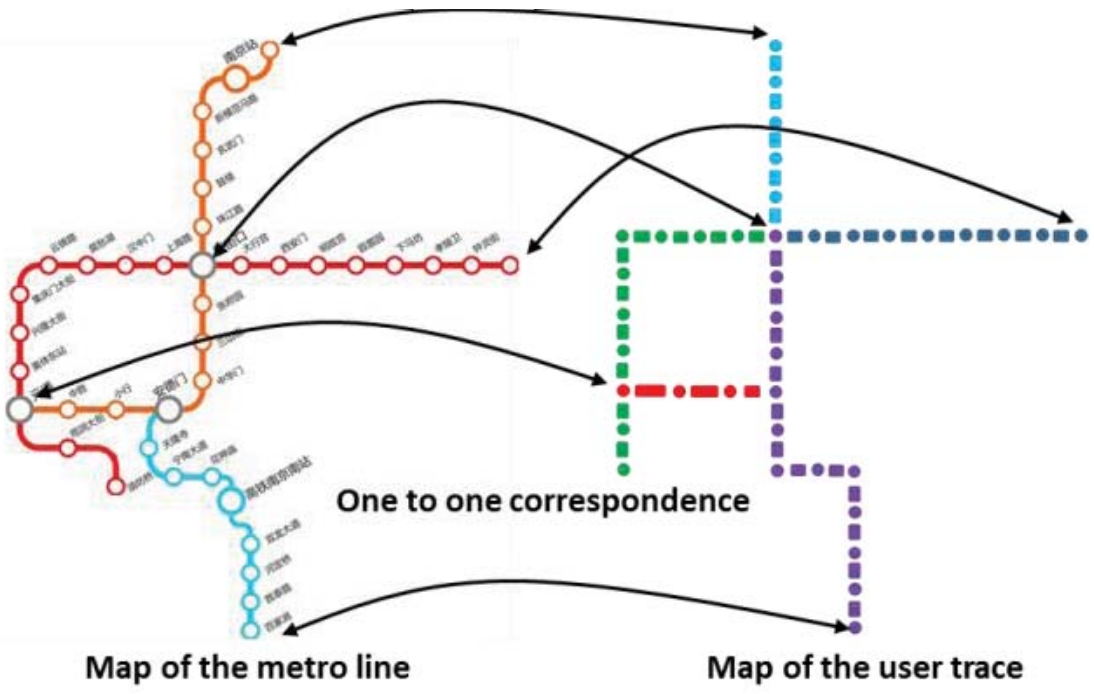

Fig. 9. Map the trace graph to the metro map

### 1) *Pattern matching based on DTW*

Fig. 10(a) and Fig. 10(b) show the barometer and magnetometer traces collected when the trains run in the same tunnel. We can see that the fluctuation of the waveforms show similar patterns. The data length is different and the waveforms are observed a shift. This is because the time cost for the trains to pass a tunnel may have little change based on the traffic. In the cloud server, we obtain the data of train running in a tunnel from the running events in user traces. Based on pattern matching, we can find the running events of the same tunnel from different users. This approach is shown as follows.

*a) Feature extraction*

First, the raw barometer and magnetometer readings may contain noise. If a data point value has an apparent spark noise, it will be removed. After removing the noise, we smooth the readings with a window of 10. In order to compare the two traces, a simple approach is to use the absolute value as the feature and calculate the mean squared error (MSE) of the two waveforms. Since the users' phones have not been calibrated (i.e., the readings of the two phones are different at the same place), there exists an unknown constant drift. This will cause

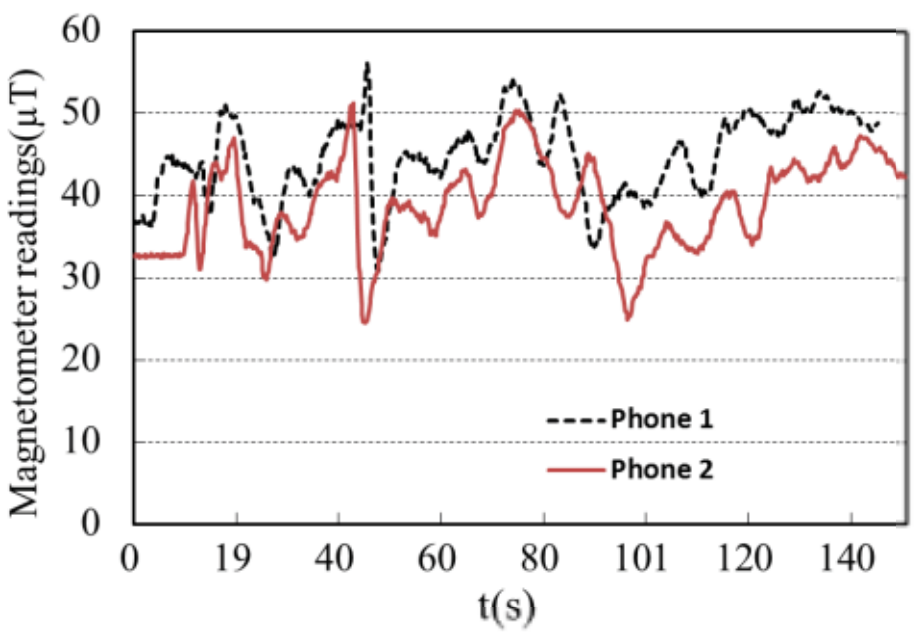

(a) The magnetometer readings for two trips in the same tunnel

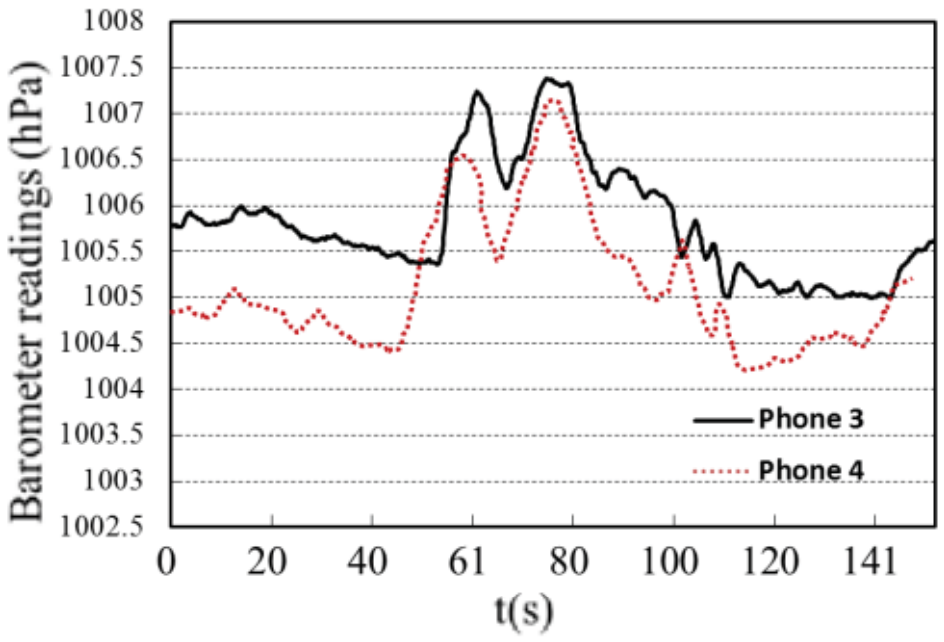

(b) The barometer readings for two trips in the same tunnel.

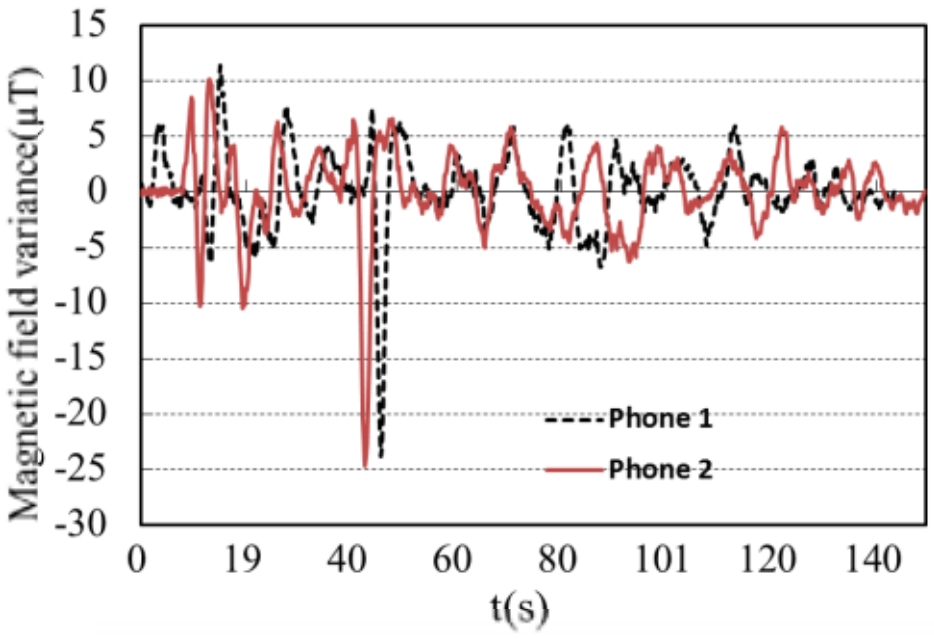

(c) The variance feature of the magnetometer readings

Fig. 10. The reading patterns of magnetometer and barometer in tunnels

error when calculating the MSE value. More importantly, the two time series has different lengths, which cannot be handled by MSE. In our approach, we use the variance as a feature to capture the fluctuation change. For both barometer and magnetometer traces, we obtain the variances as follows.

$$\mathrm{VarB(t)} = \mathrm{B(t+\Delta t)} - \mathrm{B(t)} \quad (9)$$

$$\mathrm{VarM(t)} = \mathrm{M(t+\Delta t).m} - \mathrm{M(t).m} \quad (10)$$

For example, Fig. 10(c) shows the variance of the two magnetic field traces of Fig. 10(a).

*b) Mathing with DTW*

Since each tunnel may have a different length, and the traces we collect will have different lengths of data. We apply the Dynamic Time Warping Distance Measure (DTW) [16] which is less sensitive to the time shift. To calculate the DTW, we first align the two waveforms as shown in Fig. 11. For example, for two time series of magnetic field variance $\mathrm{VarM_s}$ and $\mathrm{VarM_l}$, where

$$\mathrm{VarM_s} = s_1 s_2\, s_3\, s_4 \cdots s_n$$

$$\mathrm{VarM_l} = l_1 l_2\, l_3\, l_4 \cdots l_m$$

the sequences $\mathrm{VarM_s}$ and $\mathrm{VarM_l}$ can be arranged to form a n-by-m plane or grid, where each grid point(i, j) corresponds to an alignment between elements $s_i$ and $l_j$. A warping path, W, maps or aligns the elements of $\mathrm{VarM_s}$ and $\mathrm{VarM_l}$.

$$W = w_1 w_2\, w_3\, w_4 \cdots w_k$$

The Dynamic Time Warping distance between two time series $\mathrm{VarM_s}$ and $\mathrm{VarM_l}$ is then:

$$\mathrm{DTW}(VarM_s, VarM_l) = \partial(First(VarM_s), First(VarM_l)) + min \begin{cases} \mathrm{DTW}(VarM_s, Rest(VarM_l)) \\ \mathrm{DTW}(VarM_l, Rest(VarM_s)) \\ \mathrm{DTW}(Rest(VarM_s), Rest(VarM_l)) \end{cases} \quad (11)$$

where First(x) is the first element of x, and Rest(x) is the remainder of the time series after the First(x) has been removed, and $\partial(\mathrm{i,j}) = (s_i - l_j)^2$.

From the DTW value, we get numeric measure of the similarity between train running events. For every two running events we can obtain the DTW distance of the magnetometer reading traces and barometer reading traces.

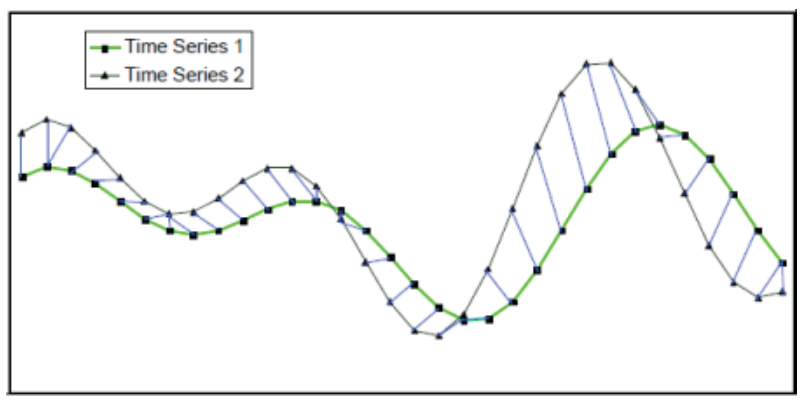

Fig. 11. Dynamic Time Warping (from [15])

*2) Merge the traces*

Traces from users often have overlaps. Given two user traces, we want to find their overlaps and merge them. Fig. 12 shows some situations where the two traces match each other. Given two traces with lengths of m and n, respectively, there are $2(m+n)$ possible overlaps. In our approach, in order to find the overlap of the two traces, we compute the average DTW value for each overlapping situation, and find the case with minimum distance. Given two user traces $\mathrm{UT_1, UT_2}$ with lengths of m and n (m<n), respectively, we get the min distance by Equation 12, where, $\mathrm{UT_i.RE_j}$ is a running event of user trace $\mathrm{UT_i}$.

$$\mathrm{Min}_{DTW} = \min \begin{cases} \mathrm{DTW(UT_1.RE_m, UT_2.RE_1)} \\ \frac{\mathrm{DTW(UT_1.RE_{m-1},UT_2.RE_1)+DTW(UT_1.RE_m,UT_2.RE_2)}}{2} \\ \cdots \\ \frac{\mathrm{DTW(UT_1.RE_1,UT_2.RE_1)+\cdots+DTW(UT_1.RE_m,UT_2.RE_m)}}{m} \\ \cdots \\ \frac{\mathrm{DTW(UT_1.RE_1,UT_2.RE_{n-m})+\cdots+DTW(UT_1.RE_m,UT_2.RE_n)}}{m} \\ \frac{\mathrm{DTW(UT_1.RE_1,UT_2.RE_{n-m+1})+\cdots+DTW(UT_1.RE_{m-1},UT_2.RE_n)}}{m-1} \\ \cdots \\ \mathrm{DTW(UT_1.RE_1, UT_2.RE_n)} \\ \mathrm{threshold} \end{cases} \quad (12)$$

If the minimum is the threshold, that means the two traces have no overlapping. If not, we conclude an overlapping case for the two traces. Then, the two traces will be merged. For example, as shown in Fig. 6(c), user C and D are matched by two overlap running events.

*3) Map train running events to tunnels*

We merge the traces from users incrementally to construct a graph, as shown in Fig. 6(c). When the graph covers all he tunnels in a metro line, it should have a one-to-one matching to the real metro map as shown in Fig. 9. Finally, in the cloud server, we obtain the patterns of all the tunnels in a metro line, and they are stored as the pattern map.

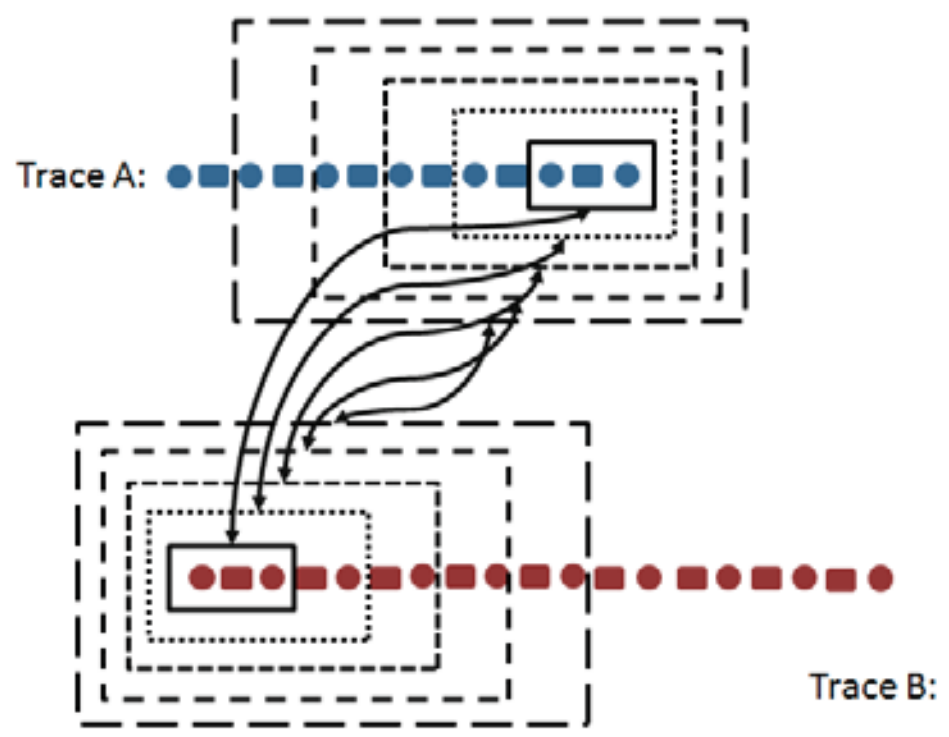

Fig. 12. The situations that two traces match each other

### C. *Locating users*

When the pattern map is generated, it can be downloaded to smartphones. M-Loc client application keeps collecting sensor data and detecting the train running events. Once detected, it finds a minimum distance tunnel by calculating the DTW distance with every tunnel in the pattern map using Equation 13. The train's position is at the end of the tunnel.

$$\mathrm{Min_{DTW}} = \min \begin{cases} \mathrm{DTW}\ (\mathrm{RE},\ \mathrm{Map.RE_1}) \\ \mathrm{DTW}\ (\mathrm{RE},\ \mathrm{Map.RE_2}) \\ \cdots \\ \mathrm{DTW}\ (\mathrm{RE},\ \mathrm{Map.RE_k}) \\ \mathrm{threshold} \end{cases} \quad (13)$$

and $\mathrm{Map.RE_k}$ is a running event of a tunnel in the pattern map, RE is the running event of a tunnel detected. If $\mathrm{Min_{DTW}}$ is the DTW distance between RE and $\mathrm{Map.RE_i}$, the position of the train now is at the end of the tunnel of $\mathrm{Map.RE_i}$.

Locating uses based on one running event only may not be accurate enough. As the train keeps running, the application will detect running events in time series. Using pattern matching based on DTW distance, we can easily map the trace in the pattern map. The matching algorithm is the same to the way we use when matching the traces in Equation 12. The accuracy grows rapidly with more and more traces collected. We get an accuracy of 97% when passenger travels 4 stations. We will show the detailed result in the next section.

## IV. EVALUATION

To evaluate M-Loc under real-world situations, we conduct a field study which involves ten users for three days in the metro lines in the city of Nanjing. The ten users are university students, among them 2 are females and 8 are males, aged between 20 and 30. Several smartphone models such as Samsung, Google Nexus, and Xiaomi are used in the experiment. Each smartphone is equipped with 3-axis accelerometer and magnetometer. Seven of them have the barometer sensor. Each smartphone is installed with M-Loc data collection software. Once started, this client software continuously collects acceleration readings and magnetic field readings at a rate of 5 samples per second (if a barometer is available, the rate is 3 samples per second). All the samples will be logged in a data file. This client software runs in the background so that the users are able to use their smartphones as usual. We conduct the field study as follows. The experiments are carried out in a Monday morning, a Thursday afternoon and a Saturday night, each lasts for three hours. We carefully choose these periods to represent different crowds (i.e., rush/non-rush hours) and weather condition (i.e., day and night). During the experiment, each user is instructed to continually take metro trains in the metro lines. The starting and ending stations are randomly chosen. To record the ground truth, microphone on smartphone was turned on to record the audio. After the experiment, we played back the audio clip to find out in which stations the users traveled. Fig. 6(a) shows part of the map of the metro line where our field study was carried out. There are 3 metro lines with 55 metro stations, 3 of them are cross-line stations.

After the experiment, we collect the logged data from each user, including the ground truth and the sensor traces. From the ground truth, we find out there are a total number of 162 one-way trips. The distribution of each trip length is shown in Fig. 13(a). It shows that there are more short trips than long trips. Using these sensor data, we run the train stop event detection algorithm. We compare the output of the detection algorithm with the ground truth. The number of stops of each trip is detected and the accuracy is shown in Table 1. The result shows that the miss detection of the train stop events rarely occurs. For the 47 trips with stations less than 4, there are only 2 trips having one missed stop event.

After the train stop event detection, we get the user trace of each user trip, including a sequence of stop and running events. Based on the ground truth, we know the running events belong to which tunnels. We randomly choose five tunnels, and use only magnetometer readings for the experiment. Fig. 13(b) shows the min, max and average DTW distance of the running events of the five tunnels. It shows that the DTW distances of running events from the same tunnel do not vary much. Fig. 14 shows the average DTW distance between the same and different tunnels. The average DTW distance of running events occurred in the same tunnel is much lower than the distance of that in different tunnels. The threshold is set to 8 in our experiment. This feature helps us to find out whether two running events occur in the same tunnel. The comparison between the DTW and MSE distance is shown in Fig. 13(c). For the reason of time drift and different lengths of the data traces, the MSE distance is much larger than the DTW distance.

Table 1

| **Missed stops** | **Missed/Total stations of trips with different length** | | |
|---|---|---|---|
| | ***<4*** | ***>4 and <12*** | ***>12*** |
| 1 | 2/47 | 3/90 | 4/21 |
| 2 | 0/47 | 2/90 | 3/21 |
| 3 | 0/47 | 0/90 | 1/21 |
| > 3 | 0/47 | 0/90 | 0/21 |

For trace matching, the result of our algorithm is shown in Fig. 13(d). There are a total of 162 traces. The left vertical axis shows the distribution of the first 300 times of trace merging. The right vertical axis shows that the matching accuracy increases when the number of overlapped stations increases. The accuracy is more than 95% when the number of overlapped stations is larger than 3. Hence, our matching algorithm only merges traces when there are more than 4 overlapped stations.

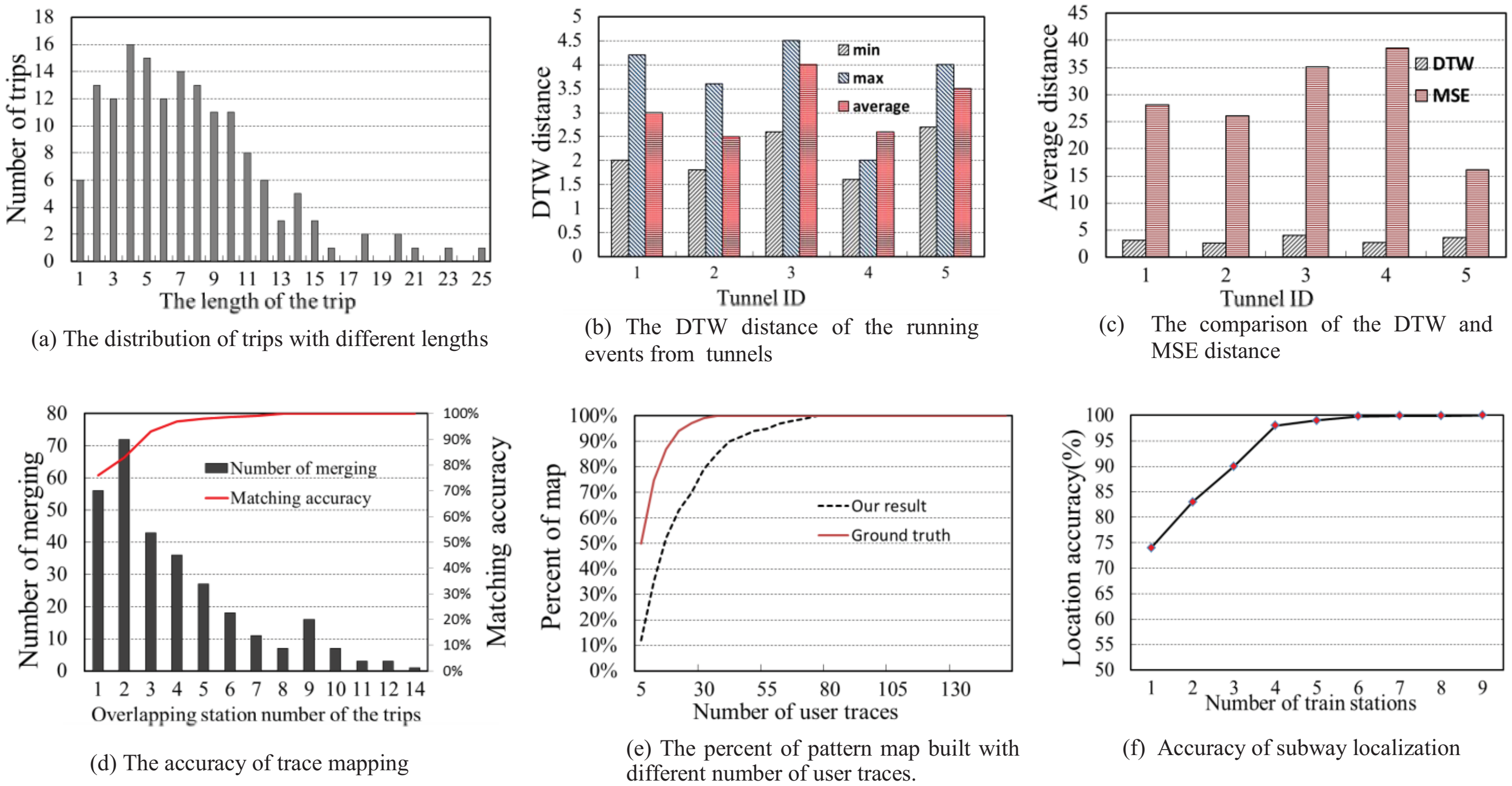

(a) The distribution of trips with different lengths

(b) The DTW distance of the running events from tunnels

(c) The comparison of the DTW and MSE distance

(d) The accuracy of trace mapping

(e) The percent of pattern map built with different number of user traces.

(f) Accuracy of subway localization

Fig. 13. The evaluation results of M-Loc

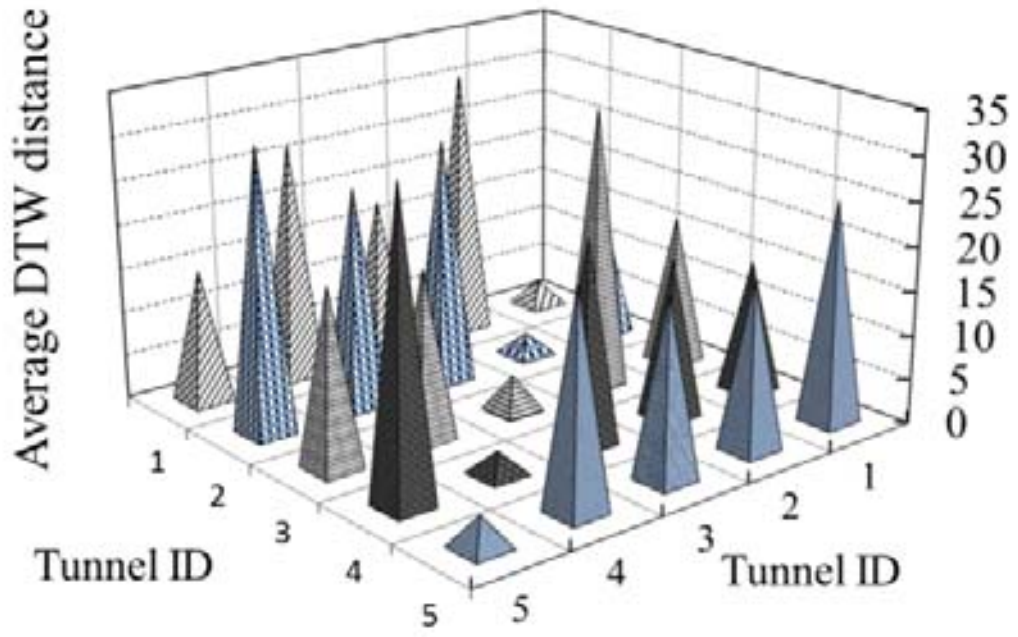

Fig. 14. The average DTW distance between running events occurred in the same and different tunnels

The merged traces form a graph. The graph grows larger when more traces are merged until it has a one-to-one mapping to the metro line map. The dotted curve in Fig. 13(e) shows the map of the metro line is successfully built with different sample size (i.e., number of user traces). The result shows that 90% of the map can be built quickly with a few user traces. The curve grows slower because some stations are not visited by passengers often, the starting and ending stations of a metro line for example. For the 3 metro lines with 55 stations, M-Loc requires only 80 user traces to build the map. The red solid line shows the ground truth of the traces to build the map. In fact, about 30 traces are required to construct the map. Our approach needs more traces because we merge two traces when there are at least 4 overlapped stations. Fig. 13(f) shows the location accuracy when the pattern map is built. When a passenger travelled only one metro station, the location accuracy is about 75%. With more stations travelled, the accuracy increases rapidly. We get an accuracy of 98% when travelling 5 stations.

In the end, we evaluate the energy consumption of M-Loc using a Samsung Galaxy Nexus smartphone running Android 4.1 OS, and the result is shown in Fig. 15. The power consumption is computed based on PowerTutor [17], a diagnostic tool for analyzing system and application power usage from the Android Market. The experiment ran for 12 hours continuously. The average power consumption of M-Loc is 109 mW. For comparison, we also show the power consumption of other localization techniques and some basic mobile functions. It shows that M-Loc consumes much less energy than the traditional localization techniques.

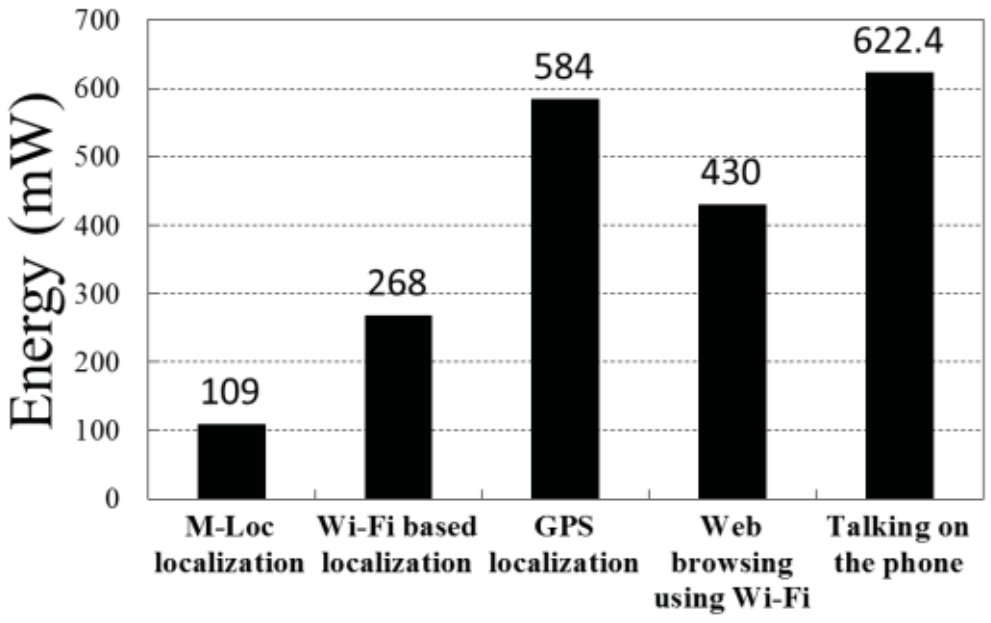

Fig. 15. The power consumption of M-Loc

## V. RELATED WORK

Many fingerprint based techniques for indoor localization have been proposed such as [6-10], which can be used to locate the user in a metro line. Existing techniques for localization rely on deployed radios (e.g., Wi-Fi access points, GSM base stations, etc) and make different assumptions about infrastructure and calibration. They mainly rely on Wi-Fi signal strength, and they are capable of achieving good accuracy in an indoor environment. Radar [6] operates on Wi-Fi fingerprints, and is capable of achieving high accuracy in indoor deployments. However, Radar needs to war-drive the entire building in order to obtain the radio map. War-driving is very time-consuming and labor-intensive. Hence, this solution is not scalable over larger areas. Some recent approaches such as LiFS [20] use crowdsourcing to reduce the training cost to some extent, but it involves a complicated training process. In reality, many mobile users may not turn on Wi-Fi all the time for energy saving, limiting the effectiveness of crowdsourcing. More importantly, deploy the Wi-Fi access points in the metro line may have security problems, which need theoretical and practice proof. The Wi-Fi infrastructure is still not widely used in today's metro lines. Different from these systems, our approach does not require any pre-installed infrastructure. It leverages on mobile phone sensing and crowdsourcing to efficiently localizing users in metro trains.

Sensor-assisted localization methods [11-14, 21] have been proposed with the popularity of smartphones, which make use of embedded sensors available on smartphones. These systems typically use accelerometer and electronic compass. However, careful calibration is needed due to the limitations of the sensing technology. Escort [11] leverages on fixed beacons for calibration and CompAcc [12] makes use of possible walking paths extracted from Google Maps [1]. We do not need war driving or calibration, and we have no assumption about users' walking patterns and the way they carry/use their smartphones. With the map of a metro line, we crowdsource sensor data from users and build the magnetic field and barometric pressure pattern map for the metro line to achieve high accuracy.

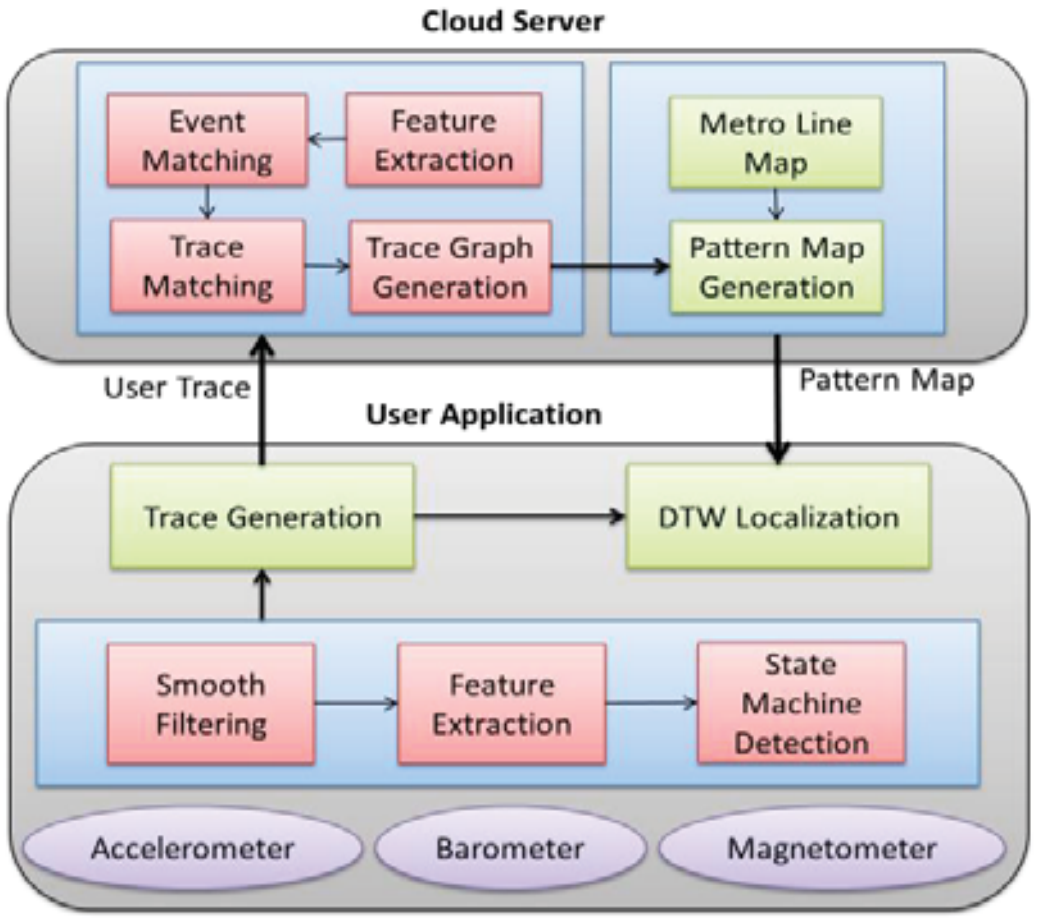

Fig. 16. Software architecture of M-Loc

## VI. CONCLUSION

This paper presents a novel, scalable metro line user localization scheme M-Loc. The software architecture of M-Loc is shown in Fig. 16. Leveraging on smartphone sensing and crowdsourcing, M-Loc requires neither any infrastructure nor war driving, making it more realistic for real-world deployment. Our field study demonstrates the performance, scalability, and robustness of M-Loc. For our future work, we will further improve M-Loc by enhancing the pattern matching algorithm. We also plan to offer a full version of M-Loc as a free service to Google's play store and the Apple store for public use, and test M-Loc under real-life situations.

## VII. ACKNOWLEDGMENT

This work was supported by the NSFC of China under Grants 61373011, 91318301and 61321491.